\documentclass[runningheads]{llncs}
\usepackage{amsmath}
\usepackage{graphicx}
\usepackage{hyperref}

\usepackage{cite}
\usepackage{subcaption}

\makeatletter
\def\thanks#1{\protected@xdef\@thanks{\@thanks
        \protect\footnotetext{#1}}}
\makeatother

\begin{document}
\title{Conditional GAN for Prediction of Glaucoma Progression with Macular Optical Coherence Tomography}
\titlerunning{cGAN based Prediction of Glaucoma Progression}
%
\author{Osama N. Hassan\inst{1}\and
Serhat Sahin\inst{2} \and
 Vahid Mohammadzadeh\inst{3}\and
 Xiaohe Yang\inst{2} \and
 Navid Amini\inst{7} \and
 Apoorva Mylavarapu\inst{5} \and
 Jack Martinyan\inst{5} \and
 Tae Hong\inst{7} \and
 Golnoush Mahmoudinezhad\inst{4} \and
 Daniel Rueckert\inst{1,6} \and
 Kouros Nouri-Mahdavi\inst{3} \and
 Fabien Scalzo\inst{2,4,8} }
\authorrunning{O. N. Hassan et al.}
%
\institute{Computing department, Imperial College London, London, UK  
\and
Department of Electrical and Computer Engineering, UCLA, Los Angeles, USA 
\and
Ophthalmology department, Jules Stein Eye Institute, Los Angeles, USA
\and
Department of Computer Science, UCLA, Los Angeles, USA 
\and
David Geffen School of Medicine, UCLA, Los Angeles, USA
\and
Technische Universität München, Munich, Germany \\
\and
Department of Computer Science, California State University, Los Angeles, USA 
\and
Department of Neurology, UCLA, Los Angeles, USA }

\thanks{O. N. Hassan and S. Sahin---Equal contribution.}

\maketitle              

\begin{abstract}
The estimation of glaucoma progression is a challenging task as the rate of disease progression varies among individuals in addition to other factors such as measurement variability and the lack of standardization in defining progression. Structural tests, such as thickness measurements of the retinal nerve fiber layer or the macula with optical coherence tomography (OCT), are able to detect anatomical changes in glaucomatous eyes. Such changes may be observed before any functional damage. In this work, we built a generative deep learning model using the conditional GAN architecture to predict glaucoma progression over time. The patient's OCT scan is predicted from three or two prior measurements. The predicted images demonstrate high similarity with the ground truth images. In addition, our results suggest that OCT scans obtained from only two prior visits may actually be sufficient to predict the next OCT scan of the patient after six months.

\end{abstract}

\section{Introduction}

Glaucoma is a progressive optic neuropathy and is the second leading cause of blindness worldwide \cite{v1}. The number of people with glaucoma worldwide was estimated to be about 60.5 million people in 2010 and it is expected to reach 111.8 million in 2040 \cite{b3}. The retinal ganglion cell (RGC) machinery is located in the inner retina and RGC axons form the optic nerve. The role of the optic nerve is to transmit visual information from the photoreceptors to the brain. Glaucoma is characterized by slow degeneration of the RGC and their axons which leads to a functional visual loss in glaucoma patients \cite{v2,v3}. The functional visual loss in glaucoma manifests as a progressive loss of vision mainly in the periphery; if glaucoma is not treated, it can eventually lead to complete visual loss and blindness \cite{b1}.   

Due to the progressive and asymptomatic nature of glaucoma, it is crucial for clinicians to diagnose it in its early stages and be able to detect its progression in a timely manner to prevent the progressive functional loss \cite{b1,b4,v4,v5,v6}. Glaucoma progression can be evaluated with structural and functional measures \cite{v7,v8,v9,v10,v11,v12,v13,v15}. The estimation of glaucoma progression is challenging as the rate of disease progression varies among individuals \cite{v16}. Moreover, measurement variability, the influence of age-related attrition, and the lack of standardization in defining progression make tracking disease deterioration a very challenging task with either structural or functional tests \cite{v17,v18,v19}. Standard achromatic perimetry and measurement of the visual field (VF) is the most common functional test used to evaluate glaucoma progression \cite{b5}. It quantifies visual degradation in the peripheral field of view of the patient. Patients may experience VF loss after a substantial amount of structural change has occurred \cite{v20}. Structural tests, such as thickness measurements of the retinal nerve fiber layer or the macula (central retina) with optical coherence tomography (OCT), are able to detect anatomical changes in glaucomatous eyes; such changes may be observed before any functional damage; hence, they may be useful for glaucoma detection especially in early stages \cite{b6,v21}. Clinicians also depend on structural tests for the detection of disease progression, especially in early to moderate stages \cite{v22,v23}. When the disease becomes more advanced, structural measurements may reach their floor and further changes might be difficult to detect \cite{v24}. At this stage, functional tests are considered to be more useful to track disease progression \cite{b7}. 

\begin{figure}[ht]
\centerline{\includegraphics[scale=0.25]{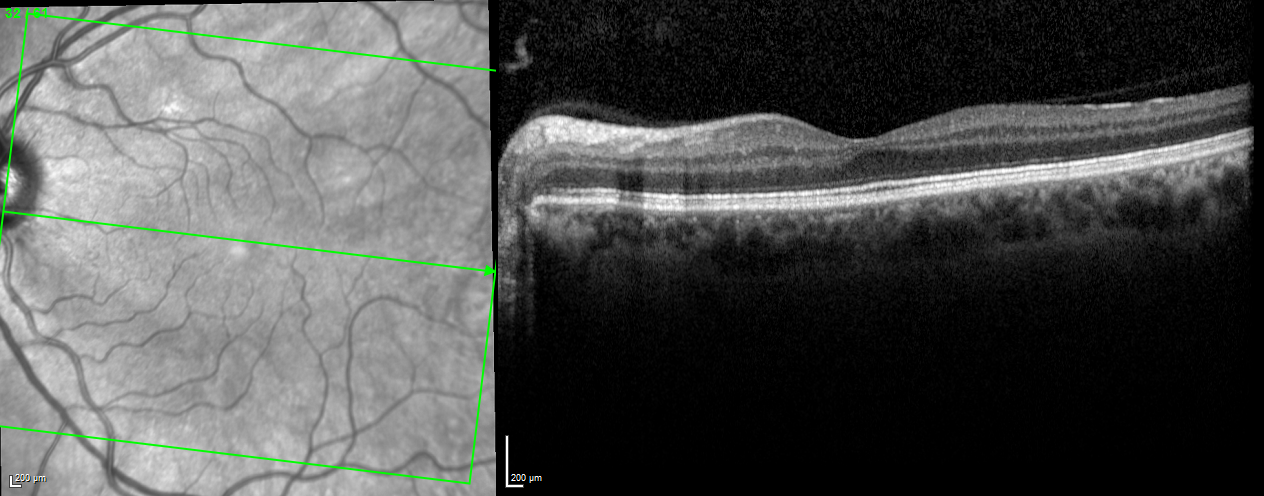}}
\caption{(Right) A raw macular B-scan of optical coherence tomography passing through the fovea (center of the macula). (Left) An infrared image of the macula and the green square outlines the area in which all the B-scans will be transmitted.}
\label{oct_example}
\end{figure}

The gold standard for retinal imaging at present is an optical imaging modality called OCT \cite{b8}. OCT is non-invasive and is able to acquire high resolution, in-vivo cross-sectional or 3D images from transparent or semi-transparent biological tissues. With the aid of OCT, it has become possible to image retinal anatomy including individual layers such as the ganglion cell layer and diagnose glaucoma before the visual field defects emerge. OCT systems can be classified into time domain based OCT (TD-OCT) and spectral domain based OCT (SD-OCT). The SD-OCT systems have better resolution, are much faster, have higher reproducibility and are more computationally efficient and therefore, SD-OCT has become the gold standard for imaging of the retinal and the optic nerve head \cite{b7}. An example of a retinal OCT cross section is shown in Fig. \ref{oct_example}.

 The goal of our work is to provide a computational framework for the modeling of glaucoma progression over time based on macular OCT images. A dataset of longitudinal macular OCT images is used. Macular OCT images of around hundred eyes with more than two years of follow-up were used. We aim to predict structural and functional changes over time. More specifically, assume we have images $x_0$, $x_i$ to $x_{n-1}$  where each image represents a scan at a specific time point $i$, the question we address here is how the image $x_n$ looks like and whether the changes are beyond what is expected. 
 A machine learning based algorithm is used to make the prediction and reconstruction of image $x_n$. Our study primarily uses the generative adversarial networks (GAN) to achieve its prediction goal. The framework of generative model is like a minimax two-player game. The GAN consists of two components: a generator $G$ and a discriminator $D$. The generator captures the data distribution and predicts the next time-point image based on the input images of previous time points. On the other hand, the discriminator tries to distinguish between the ground-truth image and the image predicted by the generator. The training succeeds when the discriminator is no longer able to tell any difference between the ground-truth images and the predicted images and the generator totally fools the discriminator. Both the generator and discriminator models are constructed using neural networks \cite{b10}.

\section{Dataset and Problem Definition}

\begin{figure}[htbp]
\centerline{\includegraphics[scale=0.30]{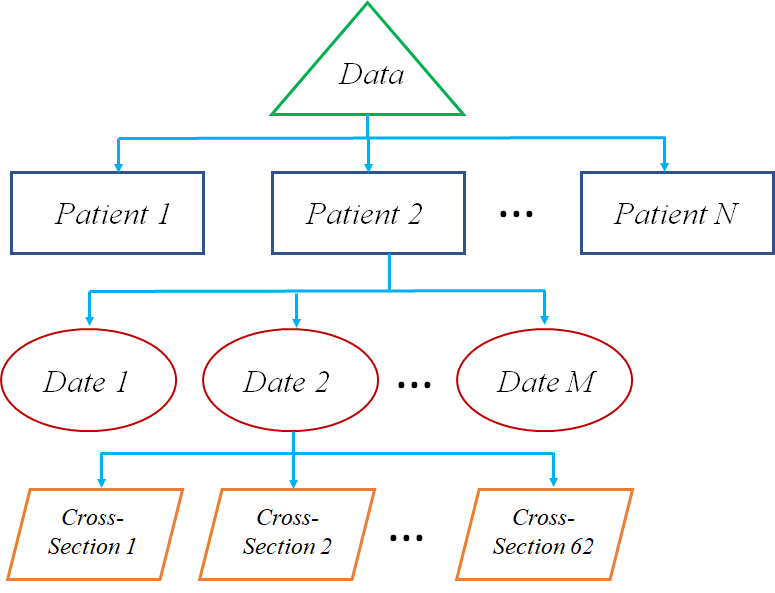}}
\caption{The hierarchy of the data set. Each patient, at each visit (date) has 61 cross-sectional images (B-scans) of the retina.}
\label{Data_structure}
\end{figure}

Our dataset consists of longitudinal macular OCT images of $109$ eyes. Each eye is scanned at four to ten visits separated by six months. Each visit has a macular OCT volume that consists of 61 cross-sectional B-scans from the central retina spanning $30X25$ degrees. The hierarchy of the dataset is depicted in Fig. \ref{Data_structure}.

The objective of this work is to predict glaucoma progression over time by the construction of a future macular cross-sectional image from past measurements. To elaborate, for a cross-sectional image that is available at $3$ time points $x_0$, $x_1$, $x_3$, we reconstruct the cross-sectional image at time point $x_{4}$. In other words, the model's task is to learn the growth of glaucoma-related features of OCT images over different cross sectional images in individual patients. Moreover, we set no constraints on the baseline which implies that the input images can be at any stage of the disease provided that subsequent images are separated by six months in time.


\section{Methods}
We adopted in our work the image-to-image translation framework with conditional generative adversarial network (cGAN) that is presented in \cite{pix2pix_paper}, with some minor modifications in our implementation of its architecture. 

\subsection{cGAN Model}
The motivation of using GAN in this problem is its flexibility of specifying the objective of the network at high-level by requiring the output of its generator to be indistinguishable from reality; the network then automatically learns the loss function that is necessary to achieve this through its adversarial mechanism. That is, the GAN learns a loss function through its discriminator that attempts to classify the generated image as true or fake while training a generative model that tries to minimize this loss at the same time. This learning-based loss function introduces a general framework to many tasks for which defining a loss function would be otherwise very difficult. In addition, we have chosen to use particularly conditional GAN to enforce the network to constrain each generated output image to the corresponding input; in other words, the output of the GAN network is conditioned on the input images \cite{pix2pix_paper}. The cGAN model consists of:

\begin{figure}[htbp]
  \centering
  \begin{subfigure}{0.8\textwidth}
  \center 
    \includegraphics[scale=0.33]{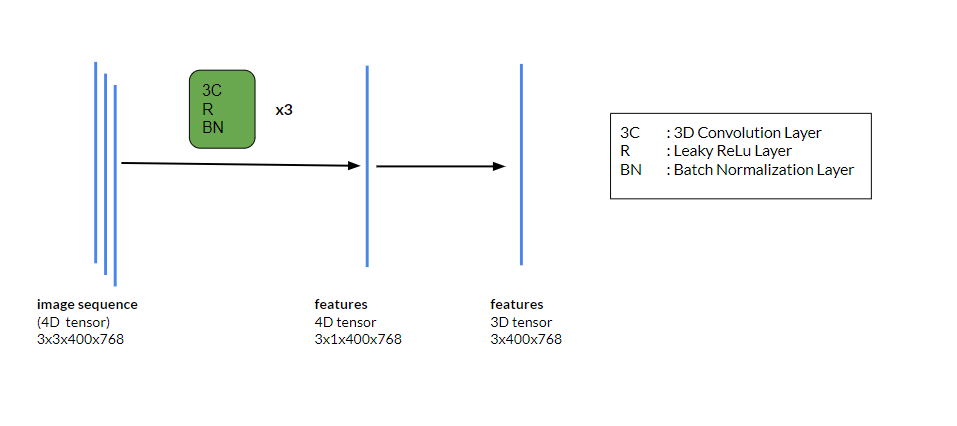}
    \caption{3D convolutional block to extract the spatio-temporal features.}
    \label{3D_CNN}
  \end{subfigure}
   \begin{subfigure}{0.8\textwidth}
   \center
    \includegraphics[scale=0.22]{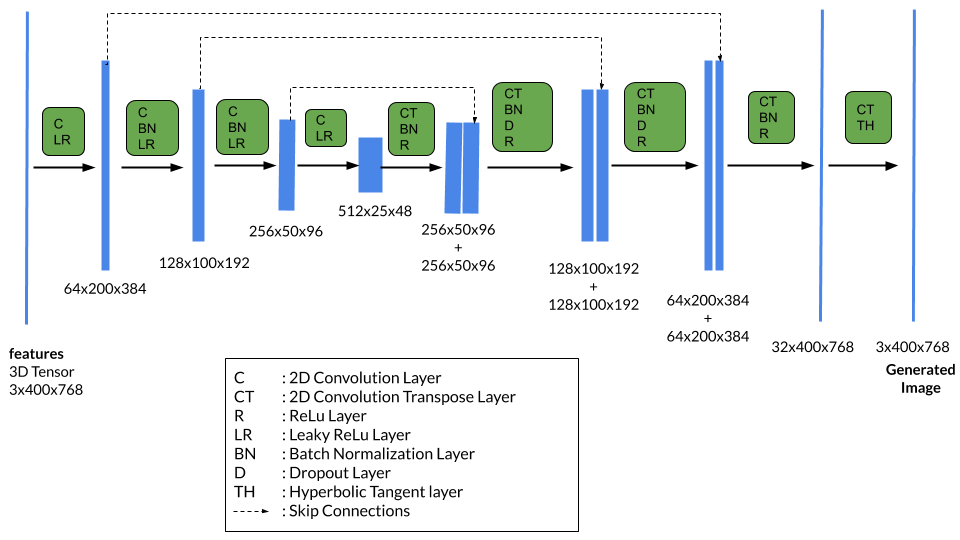}
    \caption{Generator with U-Net-based architecture. Reproduced from \cite{pix2pix_paper}.}
    \label{U_Net}
  \end{subfigure}
 
  \caption{The components of our proposed model showing the feature extraction block and the generator. }
\end{figure}

\subsubsection{Generator Model}
The generator architecture can be divided into two blocks. First, a 3D convolutional neural network (3D-CNN) block to learn the spatio-temporal features in the input image frames (see Fig.~\ref{3D_CNN}) \cite{3D_Conv1}. Second, similar to \cite{pix2pix_paper}, a U-Net based architecture, as originally proposed in \cite{unet_paper}, is used as the main block of the generative model. The general architecture is shown in Fig \ref{U_Net}.
The U-Net generator is a decoder-encoder network with long skip connections. The network consists of 4 encoding/down-sampling layers and 4 decoding/up-sampling layers. It uses a skip connection mechanism that copies the learned features from layer $i$ to layer $n-i$, where n is the total number of layers. At each layer of the generator, except for the last layer, rectified linear units (ReLU) are used in the up-sampling part of the network, and  their  leaky  version  are  used  in  the down-sampling part. In addition, batch normalization layers \cite{Batch_Normalization} are added to accelerate the training process and dropout layers are used within the up-sampling layers (except for the first and last layers) to add randomness to the generative process.


\subsubsection{Discriminator Model}
The discriminator is a fully CNN classifier. In this study, we adopted the five-layer PatchGAN discriminator that is proposed at \cite{pix2pix_paper} and originally discussed at \cite{Discr}.

\begin{figure}[htbp]
\centerline{\includegraphics[scale=0.35]{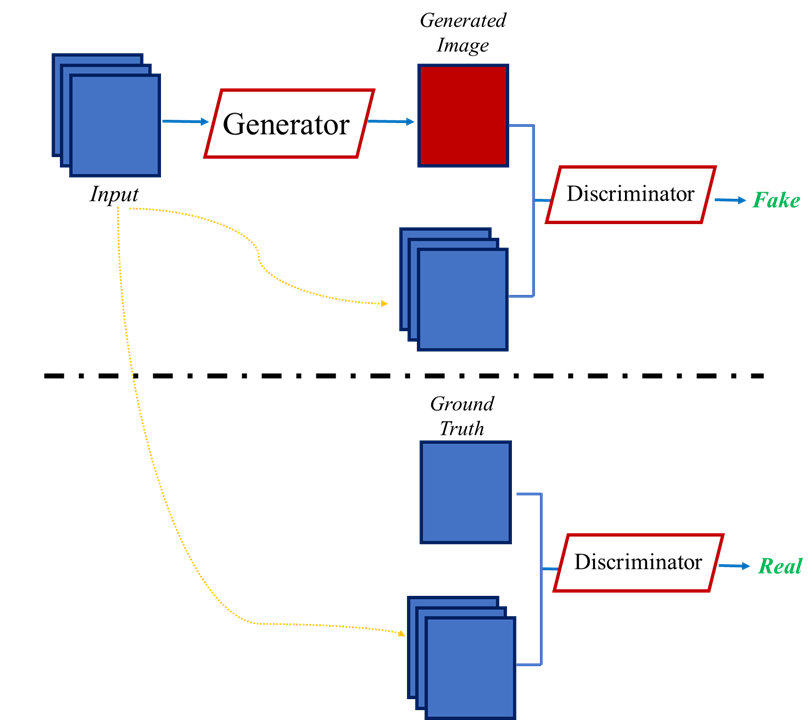}}
\caption{Training conditional GAN. (Top) the generator optimization step and (Bottom) the discriminator optimization step.}
\label{cGAN_Training}
\end{figure}

Although using the $L_1$ norm in the loss function does not preserve high frequencies and results in blurry images, it preserves the low frequency content and therefore if we use $L_1$ loss, the GAN discriminator can be designed to be more dedicated to preserving the structural and high frequency content of the generated images while leaving the the low frequency preservation task to the $L_1$ loss. In order for the discriminator to preserve high frequencies, it does not classify the image as a whole. Instead, it treats it as patches and classifies each patch as real or fake. This way the discriminator offers structured loss functionality and penalizes the joint configuration of the output and does not consider the output of each pixel to be conditionally independent in an unstructured fashion. This design modality of the discriminator is called PatchGAN since it penalizes structures at the patch scale. This results in the additional advantages of having a discriminator with fewer parameters and being able to apply the discriminator to arbitrarily large images. The PatchGAN  classifies patches of size $70x70$ as suggested by \cite{pix2pix_paper}. A concatenation of both the input images to the generator and the image to be classified are fed to the discriminator (see Fig. \ref{cGAN_Training}) and passed through five down-sampling stages resulting in a 2D map in which each pixel has a receptive field of $70x70$; the corresponding patch in the input image is then classified as real or fake. 

\begin{figure}[htbp]
    \centerline{\includegraphics[scale=0.40]{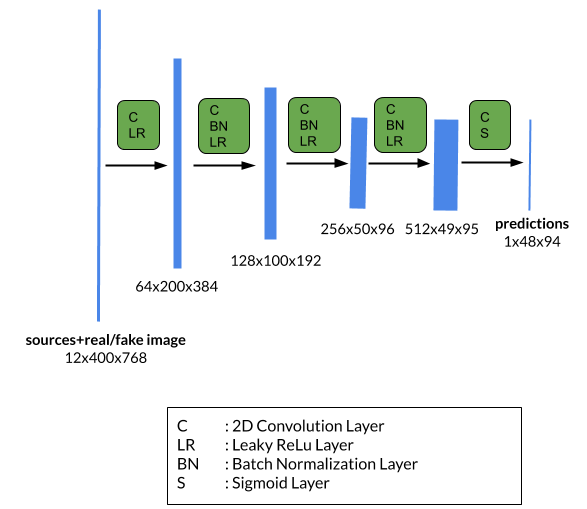}}
    \caption{PatchGAN-based discriminator Network. Reproduced from \cite{pix2pix_paper}.}
    \label{Descriminator_NET}
\end{figure}

 \subsection{Objective and Loss Functions}
 In a vanilla GAN, the generator loss ($L_G$) and the discriminator loss ($L_D$)are defined as 

 \begin{equation*}
    \begin{split}
        \label{eq:regular_losses}
         L_{G} &= F(D(\hat{y}), 1),\\
         L_{D} &=   F(D(\hat{y}), 0) + F(D( y), 1), 
     \end{split}
 \end{equation*}

 where $F$ can be a binary cross entropy (BCE) loss or mean squared error (MSE) loss, $y$ is the real ground-truth target image, and $\hat{y}$ is the predicted output of the generator. The discriminator input in this case the output of the generator $\hat{y}$. 

 However, in conditional GAN, the discriminator input includes both the generator input $x$ and the generator output $\hat{y}$. In addition, we add to the generator loss: $L_1$ norm loss, to capture the low frequency content, as explained earlier. This can be written as

 \begin{equation}
    \begin{split}
        \label{eq:pix2pix_losses}
        L_{G} &= F(D(x, \hat{y}), 1) + \alpha*L_1(\hat{y}, y), \\ 
        L_{D} &=   F(D(x, \hat{y}), 0) + F(D(x, y), 1). 
    \end{split}
 \end{equation}

 where the hyper-parameter $\alpha$ is used to emphasize the weight of the $L_1$ loss and is optimized empirically.


\section{Experiments and Results}

\subsection{Training Details}
For training, $26,592$ OCT cross-sections from 101 glaucomatous eyes were prepared. These eyes were imaged in at least four visits. We conducted two different experiments. In experiment $A$, the model was trained based on a sequence of four images using the first 3 visits as the input and the fourth visit as the output of the model. In experiment $B$, the model was trained based on a sequence of three visits using the first two visits as the input and the third visit as the output of the model. We arranged the training and validation split percentages as 75\% and 15\% respectively. The less the number of visits that one uses as an input, the more useful the model becomes when we have limited data for a given patient. 

In GAN training, it is often seen that the discriminator detects the outputs of the generator as fake images at very early stages of the training process, which stops the generator from learning. To prevent this issue, we alternated between four optimization steps on the generator and then one optimization step on the discriminator as this experimentally resulted in an optimum performance. In optimization, Adam optimizer was used for the generator with momentum parameters ${\beta}_1$ = 0.5 and  ${\beta}_2$ = 0.999. For the discriminator, stochastic gradient descent (SGD) with momentum (0.5) was used. A batch size of eight was used in training and dropout was used at rate of $0.5$ to provide noise to the GAN during training.

\subsection{Results}
The test data split had the scans of 16 patients that was not used during training the model. For the experiment A, a total of $2,379$ input-output pairs were prepared from 8 patients that have at least four visits, while in the experiment B, a total of $3,111$ input-output pairs were prepared for the all of the 16 patients' data as they all have at least three visits.
Unlike the conventional protocol, we followed \cite{pix2pix_paper} in applying dropout and batch normalization, using the test batch statistics, at the test time as well.  

\begin{figure}[htbp]
\centerline{\includegraphics[scale=0.4]{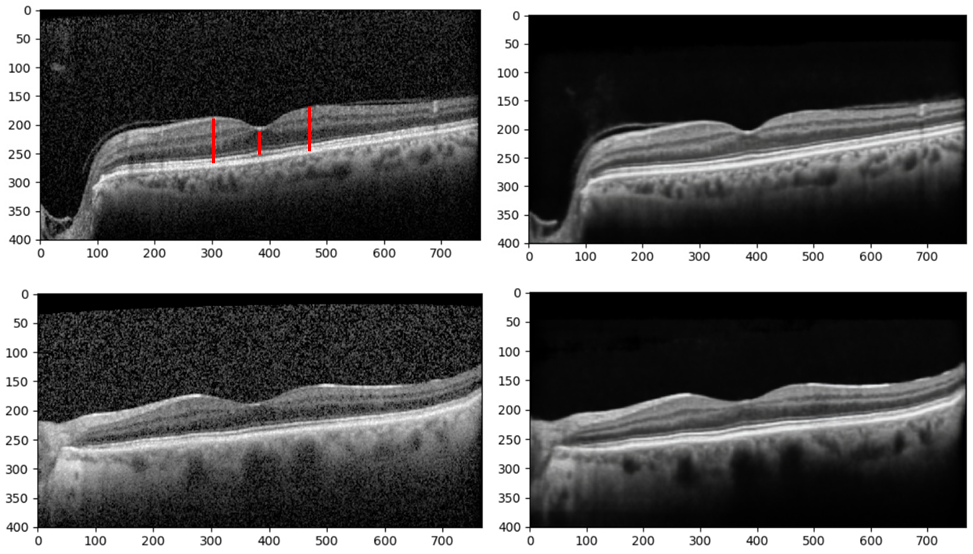}}
\caption{Examples of ground-truth macular OCT images (left column) vs. the corresponding GAN generated images (right column). The red lines highlight from left-to-right the nasal peak, the foveal pit and the temporal peak. }
\label{RealvsFake}
\end{figure}

Examples of ground-truth macular OCT images and the corresponding GAN-generated images are shown in Fig.\ref{RealvsFake}. These cross-sections pass through the fovea, which is located in the center of the macula and where the visual acuity is the best. To evaluate the accuracy of the generated images, the similarity between the original B-scans (i.e. ground truth) and the constructed B-scans (i.e. predicted B-scan) is measured by the structural similarity index measure (SSIM). SSIM takes into account changes in luminance, contrast, and structure. The SSIM index ranges in [0, 1], where 0 indicates no similarity between two images and 1 implies perfect similarity. The SSIM index has been shown to be in accordance with human visual perception and human grading of image similarity \cite{p}. Since SSIM is measured locally, it is less sensitive to noise compared with other image similarity measurements such as the mean squared error (MSE) and peak signal-to-noise ratio (PSNR)\cite{x}. Another advantage of SSIM over MSE or PNSR is that SSIM measures the perceived change in structural information, taking into account, the inter-dependencies of spatially proximate pixels and not just the error \cite{y}. The SSIM results are summarized in Table \ref{tab1}. 
 
\begin{table}[htbp]
\caption{Evaluation of the SSIM metric value for the results of experiments ``A'' and ``B''. }

\begin{center}
\centering
\begin{tabular}{ |c||c| } 
\hline
 Experiment & Average SSIM \\
 \hline  \hline 
A (with 3 visits) & 0.8325 \\
\hline 
B (with 2 visits)& 0.8336 \\
\hline 

\end{tabular}
\label{tab1}
\end{center}
\end{table}

\section{Discussion}

The visual inspection of the OCT images (i.e ground-truth) and the GAN generated images (see Fig.~\ref{RealvsFake}) initially demonstrates good structural agreement between them. Furthermore, the network has a denoising effect on the images which is evident by comparing the noise in the background of the generated and ground truth images. 

The SSIM results are above 0.83 for both experiments, which demonstrates the accuracy for our method. In addition, both experiments have very close SSIM values suggesting that it is actually adequate to use two visits to make the predictions and adding a third visit does not help the model make better predictions. This is practically very useful as it makes it possible to make predictions with limited number of visits. 

\begin{figure}[htbp]
\centerline{\includegraphics[scale=0.45]{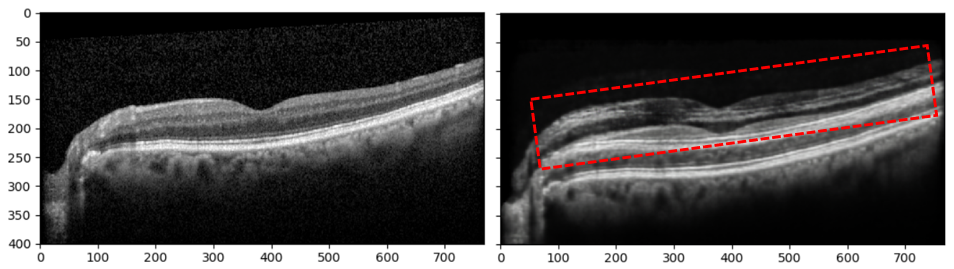}}
\caption{An example of an artifact that can be generated by the GAN network and result in a corrupted image and wrong predictions. (Left) ground truth and (Right) predicted image. Duplicate image representation can be observed.}
\label{corrupted}
\end{figure}

A limitation of our method, although uncommon, is the artifacts that can exist in the predicted image. An example of an artifact is shown in Fig.~\ref{corrupted} where the network superimposed duplicate cross-sections on top of each other. This is a weakness of the current GAN methods and represents a potential area for further research. Increasing the training dataset size or constraining the cost function with more priors or implementing a hybrid model of both learning-based and rule-based models may help us solve this problem in the future but this remains, for now, an open problem for neural networks based generative models in medical image analysis. 
 \section{Conclusion}
 Glaucoma is an eye disease that results in irreversible vision loss and is the second leading cause of blindness worldwide. Monitoring glaucoma patients for signs of progression and slowing the decay rate is the ultimate goal of glaucoma treatment. Clinicians depend on retinal structural information obtained with optical coherence tomography for tracking disease progression.
 
 In this work, we built a learning-based generative model using a conditional GAN architecture to predict glaucoma progression over time by reconstructing macular cross-sectional images from three or two prior measurements separated by six-month intervals with no constraints on the stage of the disease at the the baseline. We conducted two experiments, one with prior three visits as an input to the model and the other is only with two prior visits as the input.
 In the first experiment, a total of $2,379$ predictions were made for eight patients based on the previous three visits and the predicted images demonstrated a high similarity compared with the ground truth images with an SSIM of 0.8325. In the second experiment, a total of $3,111$ predictions were made based on two prior visits resulting in an SSIM of 0.8336. This shows that only two visits may actually be sufficient to use to make the predictions. 
 
 A limitation of our method is duplicate image artifacts that were observed in some predicted images and future work may investigate this challenge. In addition, automated segmentation based techniques that are tailored to this problem may be used as an alternative way to accurately measure the layers' thicknesses to evaluate the quality of the generated images.

%
%
%

\end{document}